\documentclass[aps,prd,groupaddress,floatfix,tighten,nofootinbib,showpacs,amsfonts,superscriptaddress]{revtex4}
\usepackage{url}
\usepackage{xcolor}
\usepackage{hyperref}

\colorlet{shadecolor}{gray!15}
\definecolor{greenLinks}{rgb}{0, 0.6, 0}
\definecolor{blueLinks}{rgb}{0, 0, 0.6}
\definecolor{redLinks}{rgb}{0.6, 0, 0}
\definecolor{tempText}{rgb}{0.55, 0.10,0.67}
\definecolor{eprintLinks}{rgb}{0.4, 0.4, 0.4}
\definecolor{journalLinks}{rgb}{0.6, 0, 0}
\newcommand{\MYhref}[3][redLinks]{\href{#2}{\color{#1}{#3}}}%
\usepackage{amssymb}
\usepackage{amsmath,color}
\usepackage{epsfig}
\usepackage{graphicx,epsf,epsfig}
\usepackage{footnote}
\usepackage{multirow}
\usepackage{color}
\def\slc#1{\setbox0=\hbox{$#1$}                  % set a box for #1
    \dimen0=\wd0                                 % and get its size
    \setbox1=\hbox{/} \dimen1=\wd1               % get size of /
    \ifdim\dimen0>\dimen1                        % #1 is bigger
       \rlap{\hbox to \dimen0{\hfil/\hfil}}      % so center / in box
       #1                                        % and print #1
    \else                                        % / is bigger
       \rlap{\hbox to \dimen1{\hfil$#1$\hfil}}   % so center #1
       /                                         % and print /
    \fi}

\def\be{\begin{equation}}
\def\ee{\end{equation}}
\def\gs{\mathrel{
   \rlap{\raise 0.511ex \hbox{$>$}}{\lower 0.511ex \hbox{$\sim$}}}}
\def\ls{\mathrel{
   \rlap{\raise 0.511ex \hbox{$<$}}{\lower 0.511ex \hbox{$\sim$}}}}

\newcommand{\onbb}{neutrinoless double beta decay }
\newcommand{\ba}{\begin{array}{c}}
\newcommand{\baz}{\begin{array}{cc}}
\newcommand{\barrr}{\begin{array}{rrr}}
\newcommand{\bad}{\begin{array}{ccc}}
\newcommand{\bav}{\begin{array}{cccc}}
\newcommand{\baf}{\begin{array}{ccccc}}
\newcommand{\bea}{\begin{equation} \begin{array}{c}}
\newcommand{\eea}{ \end{array} \end{equation}}
\newcommand{\ea}{\end{array}}

\usepackage[normalem]{ulem}
\def\lnv{lepton number violating }

\def\21{$\mathrm{SU(2)_L \otimes U(1)_Y}$ }

\newcommand {\ignore}[1]{}
\allowdisplaybreaks \allowdisplaybreaks[2]
\newcommand{\AddrDmp}{Department of Modern Physics, University of Science and Technology of China\\
  Hefei, Anhui 230026, CHINA}
\newcommand{\AddrAHEP}{AHEP Group, Institut de F\'{i}sica Corpuscular --
  C.S.I.C./Universitat de Val\`{e}ncia, Parc Cientific de Paterna.\\
  C/Catedratico Jos\'e Beltr\'an, 2 E-46980 Paterna (Val\`{e}ncia) - SPAIN}
\begin{document}
%-----------------------------------------------------------------------------
\title{Generalized $\mu-\tau$ reflection symmetry and leptonic CP violation}
%-----------------------------------------------------------------------------
\author{Peng Chen}
\email{pche@mail.ustc.edu.cn}
\affiliation{\AddrDmp}
\author{Gui-Jun Ding}
\email{dinggj@ustc.edu.cn}
\affiliation{\AddrDmp}
\author{Felix Gonzalez-Canales}
\email{felix.gonzalez@ific.uv.es}
\affiliation{\AddrAHEP}
\author{J. W. F. Valle}
\email{valle@ific.uv.es}
\homepage[URL:]{http://astroparticles.es/}
\affiliation{\AddrAHEP}
%-----------------------------------------------------------------------------

\begin{abstract}
  We propose a generalized $\mu-\tau$ reflection symmetry to constrain
  the lepton flavor mixing parameters. We obtain a new correlation
  between the atmospheric mixing angle $\theta_{23}$ and the ``Dirac''
  CP violation phase $\delta_{\rm CP}$.  Only in a specific limit our
  proposed CP transformation reduces to standard $\mu-\tau$
  reflection, for which $\theta_{23}$ and $\delta_{CP}$ are both
  maximal. The ``Majorana'' phases are predicted to lie at their
  CP-conserving values with important implications for the \onbb
  rates. We also study the phenomenological implications of our scheme
  for present and future neutrino oscillation experiments including
  T2K, NO$\nu$A and DUNE.
\end{abstract}
%-----------------------------------------------------------------------------
\maketitle
%-----------------------------------------------------------------------------
%-----------------------------------------------------------------------------
\section{Introduction}\label{sec:introduction}
%-----------------------------------------------------------------------------
The understanding of flavor mixing and CP violation is a long-standing
open question in particle physics. In order to shed light upon the
structure of fermion mixing various types of flavor symmetry-based
approaches have been
invoked~\cite{Altarelli:2010gt,Ishimori:2010au,Morisi:2012fg,King:2013eh,King:2014nza}.
Non-Abelian flavor symmetries provide a specially attractive framework.
These are typically broken spontaneously down to two distinct residual
subgroups in the neutrino and charged lepton sectors, the mismatch between
the two leading to specific lepton mixing patterns. A complete
classification of lepton mixing matrices from finite residual flavor
symmetries has been recently given in~\cite{Fonseca:2014koa}. The
precise measurement of a non-zero reactor
angle~\cite{An:2012eh,Abe:2011sj,Adamson:2013whj,Ahn:2012nd} excludes
several flavor symmetry groups and encourages future searches for CP
violation in neutrino oscillations.
It is interesting to notice that a nearly maximal CP-violating phase
$\delta_{\rm CP} \simeq 3\pi/2$ has been reported by the
T2K~\cite{Abe:2015awa}, NO$\nu$A~\cite{Bian:2015opa} and
Super-Kamiokande experiments~\cite{SK:Kachulis}, although the
statistical significance of all these experimental results is below
$3\sigma$ level. Moreover, such hints of a nonzero $\delta_{\rm CP}$ were
already present in global analyses of neutrino oscillation data, such
as the one in Ref.~\cite{Forero:2014bxa}.

Generic lepton mass matrices may admit both remnant CP symmetries as
well as remnant flavor symmetries. Moreover remnant flavor
symmetries can be generated by remnant CP
transformations~\cite{Chen:2014wxa,Chen:2015nha}. As a result it is
an interesting idea to constrain the lepton flavor mixing matrix
from CP symmetries rather than flavor symmetries. In particular, the
maximal Dirac CP-violating phase can be explained by the so-called
$\mu-\tau$ reflection symmetry under which a muon~(tau) neutrino is
transformed into a tau~(muon)
antineutrino~\cite{Harrison:2002kp,Grimus:2003yn,Farzan:2006vj}. Here
we obtain a generalized $\mu-\tau$ reflection symmetry in the
context of models based on remnant CP symmetries.

The plan of the paper is as follows. The general form of lepton
mixing is reviewed in Sec.~\ref{sec:gener-lept}. Based on the
residual CP transformation approach we derive in
Sec.~\ref{sec:generalized_mu-tau_reflection} a master formula for
the lepton mixing matrix. With this we generalize the $\mu-\tau$
reflection, and show explicitly how the CP phase can be constrained
by the experimental measurement of the atmospheric mixing angle. In
Sec.~\ref{sec:pheno} we investigate the phenomenological
implications of our scheme for current and upcoming neutrinoless
double beta decay as well as neutrino oscillation experiments.
%%
%-----------------------------------------------------------------------------
\section{General form of lepton mixing }\label{sec:gener-lept}
%-----------------------------------------------------------------------------
%%
We start with the fully ``symmetrical'' presentation of the most
general unitary lepton mixing matrix, as originally proposed in
Refs.~\cite{Schechter:1980gr,Rodejohann:2011vc}, given as:
\begin{equation}\label{eq:symmetric_para}
 {\bf U}_{\rm Sym} =
 \left( \begin{array}{ccc}
   c_{12} c_{13} &
   s_{12} c_{13} e^{ - i \phi_{12} } &
   s_{13} e^{ -i \phi_{13} } \\
  -s_{12} c_{23} e^{ i \phi_{12} }
   - c_{12} s_{13} s_{23} e^{ -i ( \phi_{23} - \phi_{13} ) } &
   c_{12} c_{23} - s_{12} s_{13} s_{23}
    e^{ -i ( \phi_{23} + \phi_{12} - \phi_{13} ) } &
   c_{13} s_{23} e^{- i \phi_{23} } \\
   s_{12} s_{23} e^{ i ( \phi_{23} + \phi_{12} ) }
    - c_{12} s_{13} c_{23} e^{ i \phi_{13} } &
   - c_{12} s_{23} e^{ i \phi_{23} }
    - s_{12} s_{13} c_{23} e^{ -i ( \phi_{12} - \phi_{13} ) } &
   c_{13} c_{23}
 \end{array} \right)\,,
\end{equation}
where $c_{ij}=\cos\theta_{ij}$ and $s_{ij}=\sin\theta_{ij}$. In this
parametrization the relation between flavor mixing angles and the
magnitudes of the entries of the leptonic mixing matrix is
\begin{equation}\label{eq:UU}
 \sin^{2} \theta_{13} = \left| U_{e3} \right|^{2} \, , \quad
 \sin^{2} \theta_{12} = \frac{ \left| U_{e2} \right|^{2} }{ 1 - \left| U_{e3} \right|^{2} }
  \quad \textrm{and} \quad
 \sin^{2} \theta_{23} = \frac{ \left| U_{\mu 3} \right|^{2} }{ 1 - \left| U_{e3} \right|^{2} } \,.
\end{equation}
The Particle Data Group presents this parametrization of the mixing
matrix in a non symmetrical form~\cite{Agashe:2014kda}, in which the
two ``Majorana'' phases appear in the diagonal (there are in principle
three ways of doing this). The resulting presentation is motivated by
the simple description of neutrino oscillation that results, in which
the ``Majorana'' phases manifestly drop out, as they
should~\footnote{Of course the Majorana phases also drop out when
  writing in the symmetric form, but in a less obvious way.}.
It is very simple to relate both presentations through a similarity
transformation involving a diagonal phase matrix~(the reader can
verify this by using Eq.~(2.5) in~\cite{Schechter:1980gr}).

First notice that the above expressions in Eq.~\eqref{eq:UU} also hold
when using the PDG form. Therefore, the difference between both
parameterizations appears only in the way of writing the CP
invariants. We start with the usual Jarlskog invariant describing CP
violation in conventional neutrino oscillations. This is defined as
$$J_{\rm CP}={\cal I} m \left \{ U_{e1}^{*} U_{\mu 3}^{*} U_{e3} U_{\mu 1} \right \},$$
and takes the following form in the symmetric parametrization
\begin{equation}\label{INV:JCP}
J_{\rm CP}=\frac{1}{8}\sin 2 \theta_{12} \, \sin 2 \theta_{23} \, \sin 2 \theta_{13}\,\cos\theta_{13}\,\sin(\phi_{13}-\phi_{23}-\phi_{12})\,.
\end{equation}
This invariant is the leptonic analogue of that which characterizes
the quark CKM mixing matrix. It is clear that, as expected, in the
symmetrical parametrization $J_{\rm CP}$ depends, apart from the three
mixing angles, on the rephasing invariant phase combination
$\phi_{13}-\phi_{23}-\phi_{12}$. This gives a very transparent
interpretation of the ``Dirac'' leptonic CP invariant. On the other
hand, concerning the remaining two invariants
$$I_{1} = {\cal I}m \left \{ U_{e2}^{2} U_{e1}^{* 2} \right \}
\quad \textrm{and} \quad I_{2} = {\cal I}m \left \{ U_{e3}^{2}
  U_{e1}^{* 2} \right \}\, ,$$ associated with the ``Majorana''
phases~\cite{Branco:1986gr,Jenkins:2007ip,Branco:2011zb} they take the
form
\begin{equation}\label{INV:I1-I2}
 \begin{array}{l}
 I_{1} = \frac{1}{4} \sin^{2} 2 \theta_{12} \cos^{4} \theta_{13} \sin ( - 2 \phi_{12} ) \quad \textrm{and} \quad
  I_{2} = \frac{1}{4} \sin^{2} 2 \theta_{13} \cos^{2}  \theta_{12} \sin ( - 2 \phi_{13} )\,.
 \end{array}
\end{equation}
These invariants appear in \lnv processes such as \onbb which do not
depend, as expected, on the ``Dirac'' invariant $J_{\rm CP}$. Indeed,
one can easily check that this is so. In contrast, however, when
written in the PDG form, the amplitude for \onbb involves all three CP
phases. Pulling out an overall phase is, of course, possible but would
bring in an ambiguity in the extraction of the phases. For all the
reasons explained in this section, we prefer the fully symmetric
parametrization to the equivalent PDG form.
\section{Generalized $\mu-\tau$ reflection}\label{sec:generalized_mu-tau_reflection}
\begin{figure}[hptb]
\begin{tabular}{cc}
 \includegraphics[width=0.4\linewidth]{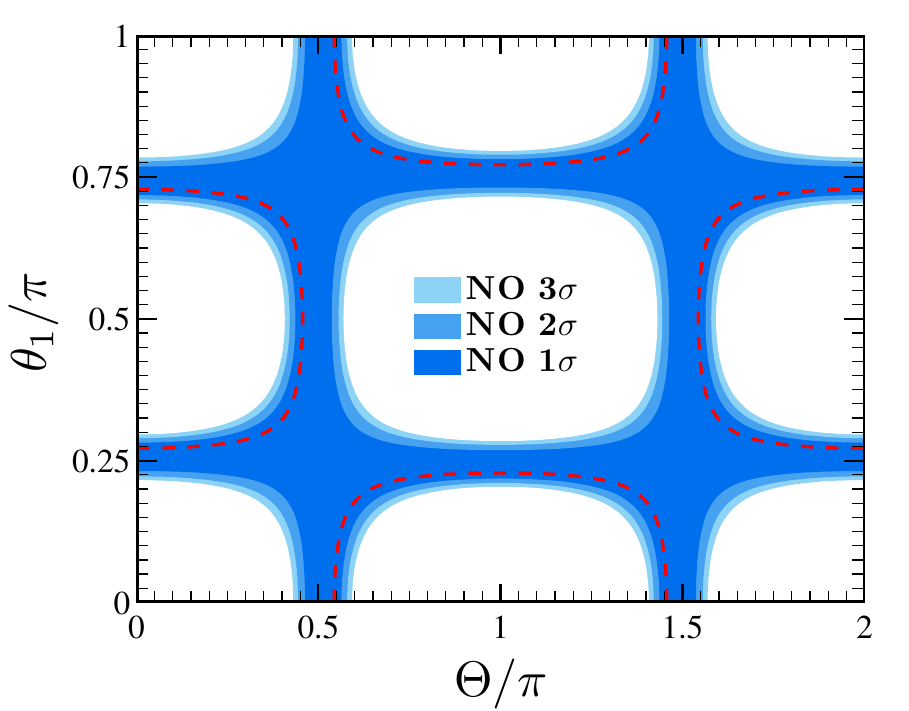} &
 \includegraphics[width=0.4\linewidth]{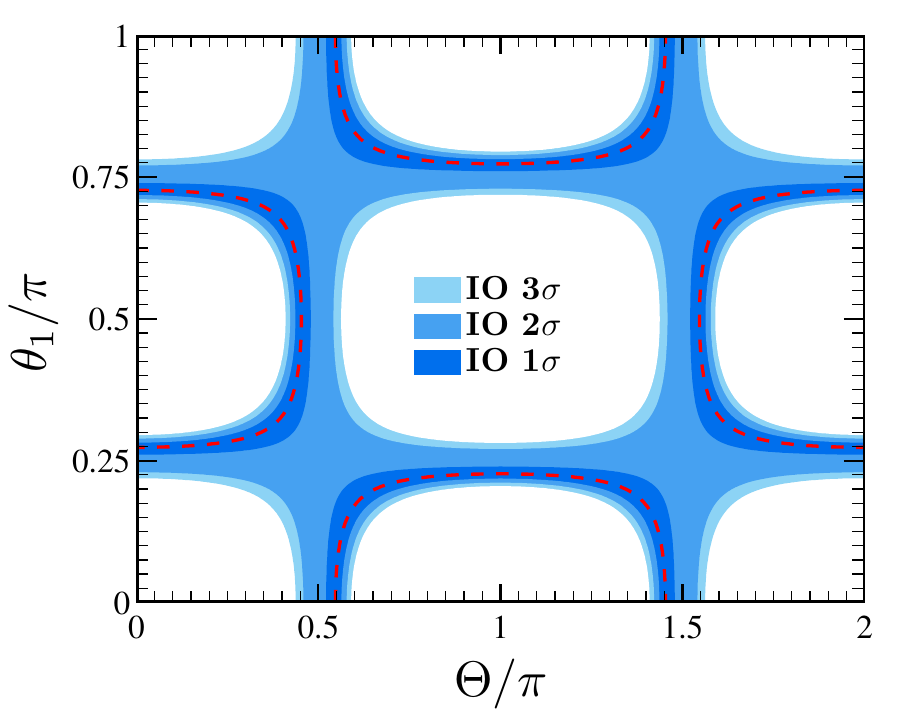}
\end{tabular}
\caption{\label{fig:contour_sinsqTheta23}
 The contour region of $\sin^2\theta_{23}$ in the plane of $\theta_{1}$ and
 $\Theta$ for both normal ordering (NO) and inverted ordering (IO) mass
 spectrum. The different contours correspond to $1\sigma$, $2\sigma$ and
 $3\sigma$. The red solid lines represent the best fitting values.}
\end{figure}

\begin{figure}[hptb]\centering
\includegraphics[width=0.6\linewidth]{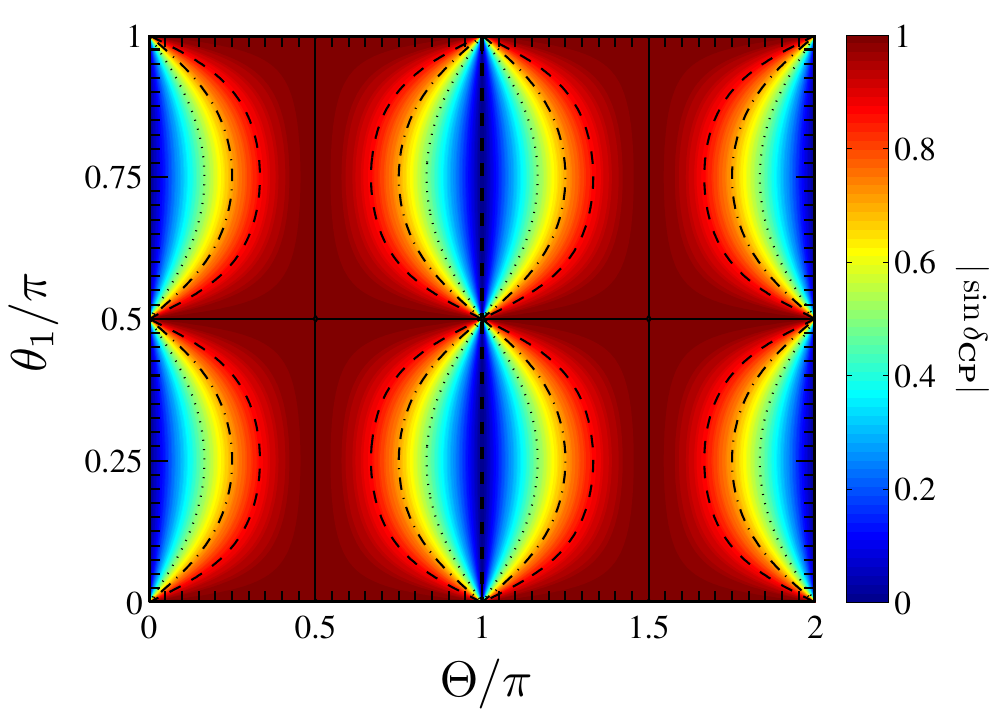}
\caption{\label{fig:contour_sinDeltaCP_abs} Contour plot of
  $|\sin\delta_{CP}|$ defined in
  Eq.~\eqref{eq:mixing_para_two_zero1b}. The thick dashed lines,
  dotted lines, dot-dashed lines, dashed line and thick solid lines
  refer to $|\sin\delta_{\rm CP}|=0$, $1/2$, $1/\sqrt{2}$, $\sqrt{3}/2$
  and 1 respectively.}
\end{figure}

In contrast with flavor symmetry schemes, our generalized CP symmetry approach can constrain not only the mixing angles but also the CP violating phases. It can lead to rather predictive scenarios, where all the mixing parameters depend on a small number of free parameters~\cite{Feruglio:2012cw}. We now turn to the method of residual CP symmetry transformations proposed in Ref.~\cite{Chen:2014wxa}. This will allow us to obtain CP-violating extensions systematically. Moreover it will, in principle, allow us to make CP predictions, starting from the general CP-conserving form of the lepton mixing matrix. Without loss of generality, we adopt the charged lepton diagonal basis, i.e.  ${\bf m}_{l} \equiv \text{diag}\left( m_{e}, m_{\mu}, m_{\tau} \right)$. Then the neutrino mass
matrix ${\bf m}_{\nu}$ can be expressed via the mixing matrix ${\bf
  U}$ as ${\bf m}_{\nu} = {\bf U}^{\ast} \text{diag} \left( m_1, m_2,
  m_3 \right) {\bf U}^{\dagger}$ under the assumption of Majorana
neutrinos.  The invariance of the neutrino mass matrix under the
action of a CP transformation ${\bf X}$ implies~\cite{Chen:2014wxa}
\begin{equation}
 {\bf X}^{T}{\bf m}_{\nu} {\bf X} = {\bf m}^{\ast}_{\nu}\,,
\end{equation}
where ${\bf X}$ should be a symmetric unitary matrix to avoid
degenerate neutrino masses. As a result we find a master formula for
the lepton mixing matrix~\cite{Chen:2014wxa}
\begin{equation}\label{eq:PMNS_single_CP}
 {\bf U}= {\bf \Sigma} \, {\bf O}_{3\times3} \, {\bf Q}_{\nu} \,,
\end{equation}
where ${\bf \Sigma}$ is the Takagi factorization matrix of ${\bf X}$
fulfilling ${\bf X}={\bf \Sigma}{\bf \Sigma}^{T}$, ${\bf Q}_{\nu}$ is
a diagonal phase matrix whose form is
{\small${\bf Q}_{\nu}=\textrm{diag}\left(e^{-ik_{1}\pi/2}, \,
    e^{-ik_{2}\pi/2}, \,e^{-ik_{3}\pi/2}\right)$} with the natural
numbers $k_{i} = 0,1,2,3$. Actually, the entries of ${\bf Q}_{\nu}$
are $\pm 1$ and $\pm i$ which encode the CP-parity or CP-signs of the
neutrino states and it renders the light neutrino mass eigenvalues
positive~\cite{Schechter:1981hw}. The matrix ${\bf O}_{3\times3} =
{\bf O}_{1} {\bf O}_{2} {\bf O}_{3}$ is a generic three dimensional
real orthogonal matrix, and it can be parameterized as
\begin{equation}\label{eq:Orthogonal_matrix}
 {\bf O}_{1} =
  \left( \begin{array}{ccc}
   1 & 0 & 0 \\
   0 &   \cos \theta_1  &  \sin \theta_1 \\
   0 & - \sin \theta_1  &  \cos \theta_1
  \end{array} \right), \,
 {\bf O}_{2} =
  \left( \begin{array}{ccc}
   \cos \theta_2  &  0  &  \sin \theta_2 \\
     0   &   1   &   0  \\
   - \sin \theta_2 &  0  &  \cos \theta_2
  \end{array} \right) \, \textrm{and} \,
  {\bf O}_{3} =
  \left(\begin{array}{ccc}
    \cos \theta_3  &  \sin \theta_3 & 0 \\
   -\sin \theta_3  &  \cos \theta_3 & 0 \\
   0  &  0  &    1
  \end{array} \right) \; .
\end{equation}
A possible overall minus sign of ${\bf O}_{3\times3}$ is dropped since
it is irrelevant. Therefore the lepton mixing matrix is predicted to
depend on three free parameters $\theta_{1, 2, 3}$ besides the
parameters characterizing the residual CP transformation ${\bf
X}$. Notice that if ${\bf \Sigma}$ is a Takagi factorization matrix of ${\bf X}$, ${\bf\Sigma}\mathbf{O}^{\prime}_{3\times3}$ is also a valid Takagi
factorization matrix, where $\mathbf{O}^{\prime}_{3\times3}$ is an arbitrary real orthogonal matrix which can be absorbed into $\mathbf{O}_{3\times3}$ by parameter redefinition. As a result, the prediction for the lepton mixing matrix in Eq.~\eqref{eq:PMNS_single_CP} remains true. Here we focus on a
generalization of the widely discussed $\mu-\tau$ reflection~\cite{Harrison:2002kp,Grimus:2003yn,Farzan:2006vj}. This
interesting CP transformation takes the following form:
\begin{equation}\label{eq:X}
{\bf X}=\left(\begin{array}{ccc}
e^{i \alpha } & 0 & 0 \\
0 & e^{i \beta } \cos \Theta &  i e^{i \frac{ ( \beta + \gamma ) }{2} } \sin \Theta \\
0 & i e^{i \frac{ ( \beta + \gamma ) }{2} } \sin \Theta & e^{i \gamma } \cos \Theta
\end{array}\right) \,,
\end{equation}
where the parameters $\alpha$, $\beta$, $\gamma$, and $\Theta$ are real. The corresponding Takagi factorization
matrix is given by
\begin{equation}
 {\bf \Sigma} =
 \left( \begin{array}{ccc}
  e^{i \frac{ \alpha }{2} } & 0 & 0 \\
  0 & e^{i \frac{ \beta }{2} } \cos \frac{\Theta}{2} &
  i e^{i \frac{ \beta }{2} } \sin \frac{\Theta}{2} \\
  0 & i e^{i \frac{ \gamma }{2} } \sin \frac{\Theta}{2} &
  e^{i \frac{ \gamma }{2} } \cos \frac{\Theta}{2}
 \end{array}   \right).
\end{equation}
As a result the resulting lepton mixing angles are determined as
\begin{equation}\label{eq:mixing_para_two_zero1a}
\begin{array}{l}
\sin^{2} \theta_{13} = \sin^{2} \theta_{2}, \quad
\sin^{2} \theta_{12} = \sin^{2} \theta_{3}, \quad
\sin^{2} \theta_{23} = \frac{1}{2} \left( 1 - \cos \Theta \cos 2 \theta_{1} \right)\,,
\end{array}
\end{equation}
while the CP violation parameters are predicted as
\begin{equation}\label{eq:mixing_para_two_zero1b}
 \begin{array}{l}\vspace{2mm}
  J_{\rm CP } =
   \frac{1}{4} \sin \Theta \sin \theta_{2} \sin 2 \theta_{3} \cos^{2}\theta_2 \,, \quad
   \sin\delta_{\rm CP}=\frac{ \sin \Theta \; {\rm sign} \left[ \sin \theta_{2} \sin 2 \theta_{3} \right]
  }{
  \sqrt{ 1 - \cos^{2} \Theta \cos^{2} 2 \theta_{1}  } } \, ,\\ \vspace{2mm}
  \tan\delta_{\rm CP}=\tan\Theta\csc2\theta_{1}\,, \quad
  \phi_{12}=\frac{k_{2}-k_{1}}{2}\pi\,, \quad
  \phi_{13}=\frac{k_{3}-k_{1}}{2}\pi\,, \quad \delta_{\rm CP}=\frac{k_{3}-k_{2}}{2}\pi-\phi_{23}\,.
 \end{array}
\end{equation}
In general, as we saw in the previous section, the lepton mixing
matrix is specified by six parameters, three angles and three
phases. In our scenario only four free independent parameters appear:
$\theta_{1}$, $\theta_{2}$, $\theta_{3}$ and $\Theta$. Notice also
that the parameters $\alpha$, $\beta$ and $\gamma$ in Eq.~\eqref{eq:X}
do not appear in the mixing parameters. It follows that the three mixing angles are not correlated with each other. Hence we have no genuine prediction for mixing angles.  In contrast, however, an important prediction concerning CP violation is that the ``Majorana'' phases $\phi_{12}$ and $\phi_{13}$ are restricted to lie at their
CP-conserving values, and correspond simply to the CP parities of the
neutrino states~\cite{wolfenstein:1981rk,Schechter:1981hw}. Moreover,
one sees that the atmospheric angle and the Dirac phase $\delta_{\rm CP}$ are given in terms of two parameters $\theta_{1}$ and $\Theta$, and they are correlated with each other according to~\footnote{We note
that in the $A_4$ flavor-symmetry-based model in Ref.~\cite{Morisi:2013qna} we also have a correlation between $\delta_{\rm CP}$ and the atmospheric angle.}
\begin{equation}\label{eq:corre_theta_{23}_deltaCP}
 \sin^{2} \delta_{\rm CP} \sin^22\theta_{23}= \sin^{2} \Theta \,.
\end{equation}
Taking $\Theta=\pm\frac{\pi}{2}$, both $\theta_{23}$ and $\delta_{\rm
  CP}$ are maximal, since the residual CP transformation ${\bf X}$
reduces to the standard $\mu-\tau$ reflection. When
$\theta_{1}=\pm\frac{\pi}{4}$, the atmospheric mixing angle
$\theta_{23}$ is maximal and $\tan \delta_{\rm CP} = \pm \tan
\Theta$. On the other hand, we have maximal $\delta_{\rm CP}$ and
$\sin^{2} \theta_{23} = \sin^{2}\frac{\Theta}{2}$ for $\theta_{1}=0,
\pi$. Present global fits of neutrino oscillation data indicate the
$\theta_{23}$ deviates from the maximal
value~\cite{Forero:2014bxa}. If non-maximal $\theta_{23}$ was
confirmed by forthcoming more sensitive experiments, the standard
$\mu-\tau$ reflection would be disfavored, while our present CP
transformation would provide a good alternative, with the value of
$\Theta$ determined from the measured values of $\theta_{23}$ and
$\delta_{\rm CP}$. We display the contour regions for $\sin^{2}
\theta_{23}$ and $\left| \sin\delta_{\rm CP} \right|$ in the plane
$\theta_{1}$ versus $\Theta$ in Fig.~\ref{fig:contour_sinsqTheta23}
and Fig.~\ref{fig:contour_sinDeltaCP_abs} respectively.

\begin{figure}[hptb]\centering
 \begin{tabular}{cc}
  \includegraphics[width=0.4\linewidth]{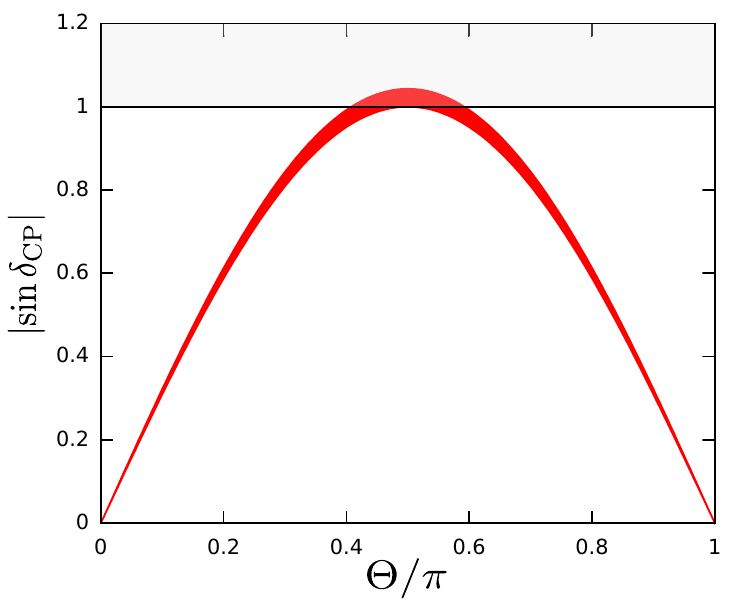}
 \end{tabular}
 \caption{The regions of $|\sin\delta_{CP}|$ versus $\Theta$, where
   the atmospheric mixing varies within its experimentally allowed
   $3\sigma$ range~\cite{Forero:2014bxa}.}\label{fig:sinDeltaCP_Theta}
\end{figure}

Given the $3\sigma$ range of the atmospheric mixing angle
$0.393\leq\sin^2\theta_{23}\leq0.643$, the correlation in
Eq.~\eqref{eq:corre_theta_{23}_deltaCP} allows us to predict the range
of the Dirac CP violating phase $| \sin \delta_{\rm CP} |$ as a
function of the parameter $\Theta$ which characterizes the CP
transformation ${\bf X}$.  The result is shown in
Fig.~\ref{fig:sinDeltaCP_Theta}.
It is remarkable that $|\sin \delta_{\rm CP} |$ is predicted to lie in
a rather narrow region for a given value of $\Theta$.
\begin{figure}[hptb]
\begin{tabular}{cc}
\includegraphics[height=7cm,width=0.4\linewidth]{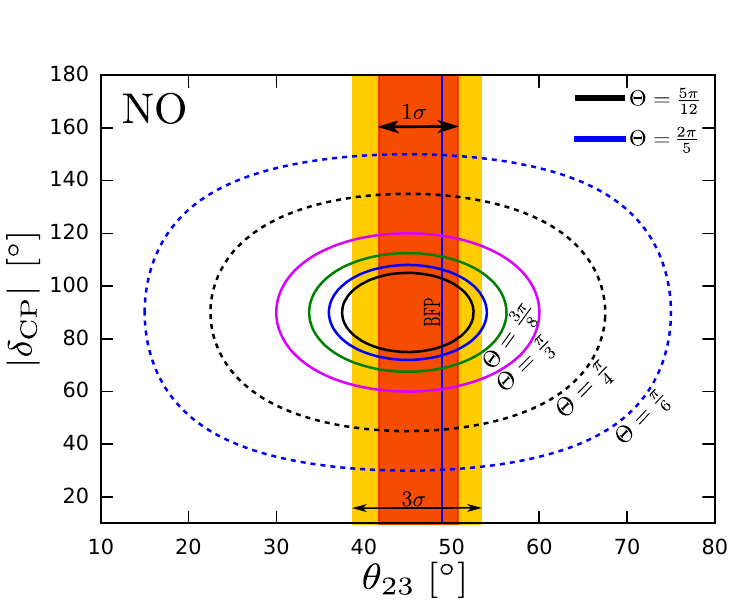} &
\includegraphics[height=7cm,width=0.4\linewidth]{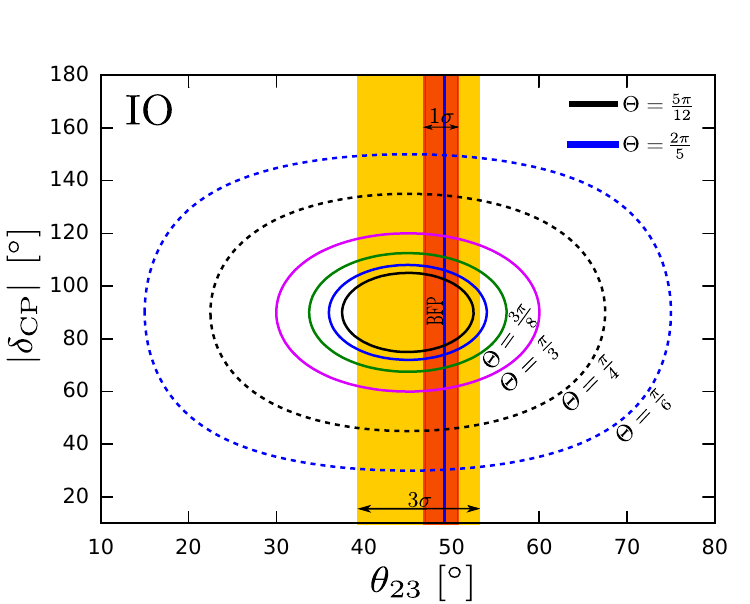}
\end{tabular}
\caption{Predicted range of $|\delta_{\rm CP}|$ phase, for given
  illustrative values of the $\Theta$ parameter characterizing our CP
  scheme, where $\Theta$ is fixed to $\pi/6$, $\pi/4$, $\pi/3$, $3\pi/8$, and
  $5\pi/12$, $2\pi/5$. The best fits, $1\sigma$ and 3$\sigma$ ranges
  of the atmospheric mixing angle from~\cite{Forero:2014bxa} are
  indicated.}\label{fig:deltaCP_Theta}
\end{figure}

On the other hand, as we can see from
Eq.~\eqref{eq:corre_theta_{23}_deltaCP}, the correlation between the
atmospheric angle and the CP phase is weighted by the value of the
$\Theta$ angle. In Fig.~\ref{fig:deltaCP_Theta} we map the allowed
ranges of the $\delta_{\rm CP}$ phase versus the atmospheric angle for
given values of the $\Theta$ parameter determining a given CP
scheme. The best fit points (BFP), $1\sigma$ and 3$\sigma$ ranges of
$\theta_{23}$ reported in~\cite{Forero:2014bxa} are indicated. For the
benchmark value of $\Theta = 3\pi/8$, $2 \pi/5$ and $5\pi/12$, the
range of $|\sin \delta_{\rm CP}|$ allowed by the data of $\theta_{23}$
at $3\sigma$ level is given in Table~\ref{tab:sin_delta}. One sees
that the experimentally observed nearly maximal $\delta_{\rm CP}$ can be
reproduced.
\begin{table}[hptb]
 \begin{center}
  \begin{tabular}{|c|c|c|c|}\hline\hline
   $\Theta$ & $3\pi/8$ & $2\pi/5$ & $5\pi/12$ \\ \hline
   $|\sin\delta_{\rm CP}|$ & $\left[ 0.92,0.96 \right]$ & $\left[ 0.95,0.99 \right]$ &
   $\left[ 0.97,1 \right]$ \\\hline\hline
  \end{tabular}
 \end{center}\renewcommand{\arraystretch}{1.0}
 \caption{\label{tab:sin_delta}Predicted range of $|\sin \delta_{\rm CP}|$ for the benchmark values
   $\Theta = 3\pi/8$, $2\pi/5$ and $5\pi/12$, allowed by the current $3\sigma$ range
   $38.8^{\circ} \leq \theta_{23} \leq 53.3^{\circ}$~given in~\cite{Forero:2014bxa}.}
\end{table}
%-----------------------------------------------------------------------------
\section{Phenomenological implications}\label{sec:pheno}
%-----------------------------------------------------------------------------
We have seen that our generalized $\mu-\tau$ reflection symmetry
schemes make well-defined predictions for CP violation. In the
following, we shall investigate the phenomenological implications of
these predictions for \lnv processes such as neutrinoless double beta
decay ($0 \nu \beta \beta$), as well as conventional neutrino
oscillations.
%
%-----------------------------------------------------------------------------
\subsection{Neutrinoless double beta decay}\label{sec:onbb}
%-----------------------------------------------------------------------------
%
The rare decay $\left(A, Z\right) \to \left( A, Z+2 \right) + e^{-} +
e^{-}$ is the \lnv process ``par excellence''. Its observation would
establish the Majorana nature of neutrinos irrespective of their
underlying mass generation
mechanism~\cite{Schechter:1981bd,Duerr:2011zd}. Within the simplest
light neutrino exchange mechanism its amplitude is sensitive to the
``Majorana phases''. Up to nuclear matrix
elements~\cite{Simkovic:2012hq} and experimental
factors~\cite{avignone:2007fu,Barabash:2011fn} the amplitude for the
decay is proportional to the effective mass parameter
\begin{equation}
 \left| m_{ee} \right| =
  \left| m_{1} \cos^{2} \theta_{12} \cos^{2} \theta_{13}
   + m_{2} \sin^{2} \theta_{12} \cos^{2} \theta_{13} e^{ - i 2 \phi_{12} }
   + m_{3} \sin^{2} \theta_{13} e^{ - i 2 \phi_{13} } \right|\, ,
\end{equation}
where we used the symmetric parametrization of the lepton mixing
matrix. It is clear that only the two ``Majorana phases'' appear but
not the ``Dirac phase''~\cite{Rodejohann:2011vc}.

The crucial prediction of our CP scheme concerns CP violation, in
particular, the absence of Majorana CP violation, as seen in
Eq.~\eqref{eq:mixing_para_two_zero1b}. Within our scheme the Majorana
phases are predicted as $\phi_{12}=\frac{k_{2}-k_{1}}{2}\pi$ and
$\phi_{13} = \frac{k_{3}-k_{1}}{2}\pi$. In other words, these phase
factors are predicted to lie at their CP conserving values, which
correspond to the CP signs of neutrino
states~\cite{wolfenstein:1981rk,Schechter:1981hw}. This implies that
the two Majorana phases $(\phi_{12}, \phi_{13})$ can only take
 the following nine values $(0,0)$, $(0,\pm \pi/2)$, $(\pm
\pi/2, 0)$ and $(\pm \pi/2, \pm \pi/2)$.

\begin{figure}[hptb]\centering
\begin{tabular}{cc}
\includegraphics[width=0.6\linewidth]{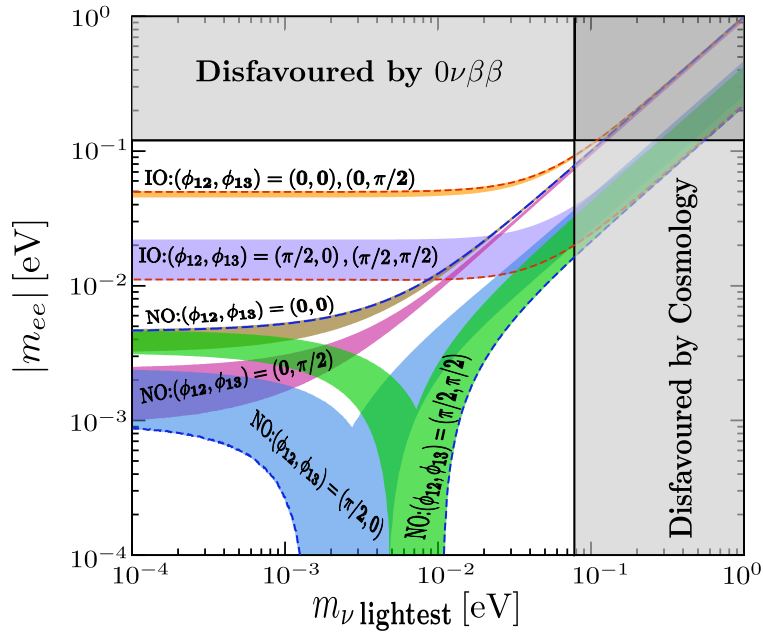}
\end{tabular}
\caption{ Effective mass $\left| m_{ee}
  \right|$ describing neutrinoless double beta decay in our scenario
  where the Majorana phases are predicted at their CP conserving
  values 0 and $\pm \pi/2$. The red and blue dashed lines indicate the
  regions currently allowed at 3$\sigma$ by neutrino oscillation
  data~\cite{Forero:2014bxa} for inverted and normal neutrino mass
  ordering, respectively. The allowed values of $\left| m_{ee}
  \right|$ for different values of $\phi_{12}$ and $\phi_{13}$ are
  displayed. For comparison we show the most stringent upper bound
  $\left| m_{ee} \right| < 0.120$eV from
  EXO-200~\cite{Auger:2012ar,Albert:2014awa} in combination with
  KamLAND-ZEN~\cite{Gando:2012zm}. The upper limit on the mass of the
  lightest neutrino is derived from the lastest Planck result
  $\sum_{i} m_{i} < 0.230$eV at $95\%$
  level~\cite{Planck:2015xua}.}\label{Fig:zbbb}
\end{figure}

The effective mass $m_{ee}$ is an even function of the phases
$\phi_{12}$ and $\phi_{13}$. Hence, the difference of signs between
Majorana phase values is irrelevant, hence the only relevant values
for Majorana phases are $(0,0)$, $(0, \pi/2)$, $(\pi/2, 0)$ and
$(\pi/2,\pi/2)$. This means that for each possible neutrino mass
ordering, there are only four independent regions for the effective
mass. Now, inputting the experimentally allowed 3$\sigma$ ranges of
neutrino oscillation parameters~\cite{Forero:2014bxa}, the resulting
regions of the effective mass $|m_{ee}|$ correlate with the lightest
neutrino mass as shown in Fig.~\ref{Fig:zbbb}.

The first comment is that, compared with the generic case, the
predictions of our scheme for the neutrino-mass-induced~\onbb
amplitude are in some cases rather powerful. Consider, for example,
the case of inverted ordering (IO), when the lightest neutrino mass is
$m_{3}$. In this case the predicted effective mass for $\phi_{13} = 0$
and $\phi_{13}=\pi /2$ almost coincide, as shown in
Fig~\ref{Fig:zbbb}. However, the predictions for $\phi_{12} = 0$ and
$\phi_{12} = \pi /2$ can be probably be distinguished from each other
in the next generation of experiments.

Turning to the case of normal neutrino mass ordering (NO) it is
remarkable that one can place a lower bound for the effective mass
despite the possibility of destructive interference amongst the three
light neutrinos.  Indeed no such interference can take place for $(0,
0)$ and $(0, \pi/2)$.  This situation is analogous to what occurs in a
number of flavour symmetry
models~\cite{Dorame:2011eb,Dorame:2012zv,King:2013hj,Bonilla:2014xla,Agostini:2015dna,Li:2014eia,Ding:2014ora,Li:2015jxa}.

For completeness we now summarize the above results as
tables~\ref{tab:dbd-cps-nh} and \ref{tab:dbd-cps-ih}, for the cases of
normal and inverted ordering, respectively. In these tables, the first
column gives possible forms of the $Q_{\nu}$ matrix, while the second
and third columns show the corresponding (CP conserving) values of the
Majorana phases, and the resulting allowed ranges for the effective
mass parameter $|m_{ee}|$.

%%
%%%%%%%%%%%%%% Mee NO %%%%%%%%%%
%%
\begin{table}[hptb]
\centering
\begin{tabular}{|c|c|c|} \hline \hline
\multicolumn{3}{|c|}{  {\footnotesize Normal Ordering} } \\ \hline \hline
{\footnotesize CP signs $Q_{\nu}$} & {\footnotesize $\left( \phi_{12}, \, \phi_{13} \right)$} &  {\footnotesize $\left| m_{ee} \right| \left( 10^{-2}\textrm{ eV} \right)$} \\ \hline
{\footnotesize diag$\left( 1, 1, 1 \right)$}  &
{\footnotesize $\left( 0, \, 0 \right)$} &
{\footnotesize $\left[\, 0.32 \, , 7.22 \, \right]$} \\ \hline
{\footnotesize diag$\left( 1, 1, -i \right)$} &
{\footnotesize $\left( 0, \, \frac{\pi}{2} \right)$} &
{\footnotesize $\left[ 9.50 \times 10^{-2} \, , 6.89 \right] $}  \\ \hline
{\footnotesize diag$\left( 1, -i, 1 \right)$} &
{\footnotesize$\left( \frac{\pi}{2} , \, 0  \right)$ } &
{\footnotesize $\left[ 0 \, , 3.31  \right]$} \\ \hline
{\footnotesize diag$\left( 1, -i, -i \right)$} &
{\footnotesize$\left( \frac{\pi}{2} , \, \frac{\pi}{2}  \right)$ } &
{\footnotesize $\left[ 0 \, , 2.94  \right] $}  \\ \hline \hline
\end{tabular}
\caption{The allowed ranges for the effective mass in \onbb for
the case of normal ordering. Notice that in our generalized $\mu-\tau$
reflection scenario the Majorana phases can only be 0 and $\pm \pi/2$.}
\label{tab:dbd-cps-nh}
\end{table}
%%
%%%%%%%%%%%%%% Mee IO %%%%%%%%%%
%%
\begin{table}
  \centering
  \begin{tabular}{|c|c|c|} \hline \hline
  \multicolumn{3}{|c|}{  {\footnotesize Inverted Ordering} } \\ \hline \hline
  {\footnotesize CP signs $Q_{\nu}$} &
   {\footnotesize $\left( \phi_{12}, \, \phi_{13} \right)$} &
   {\footnotesize $\left| m_{ee} \right| \left( 10^{-2}\textrm{ eV} \right)$} \\ \hline
  {\footnotesize $\begin{array}{c}
   \textrm{diag}\left( 1, 1, 1 \right) \\
   \textrm{diag}\left( 1, 1, -i \right)
   \end{array}$}  &
   {\footnotesize $\begin{array}{c}
   \left( 0, \, 0 \right) \\
   \left( 0, \, \frac{\pi}{2} \right)
   \end{array}$} &
   {\footnotesize $\left[\, 4.59 \, , 8.20 \, \right]$} \\ \hline
  {\footnotesize $\begin{array}{c}
   \textrm{diag}\left( 1, -i, 1 \right) \\
   \textrm{diag}\left( 1, -i, -i \right)
   \end{array}$} &
   {\footnotesize $\begin{array}{c}
   \left( \frac{\pi}{2} , \, 0 \right) \\
   \left( \frac{\pi}{2}, \, \frac{\pi}{2} \right)
   \end{array}$}  &
   {\footnotesize $\left[ 1.10 \, , 3.45 \right] $}  \\ \hline \hline
  \end{tabular}
  \caption{Same as above  for the case of inverted ordering.}
  \label{tab:dbd-cps-ih}
\end{table}
%
%-----------------------------------------------------------------------------
\subsection{CP violation in conventional neutrino oscillations}\label{sec:cp-viol-conv}
%-----------------------------------------------------------------------------
%
The existence of leptonic CP violation would show up as the difference
of oscillation probabilities between neutrino and anti-neutrinos in
the vacuum~\cite{Nunokawa2008338}:
$$\Delta P_{\alpha \beta} \equiv P
\left( \nu_{\alpha} \to \nu_{ \beta } \right) - P \left(
  \bar{\nu}_{\alpha} \to \bar{\nu}_{ \beta } \right) = - 16 \,
J_{\alpha \beta} \, \sin \Delta_{ 21 } \sin \Delta_{ 23 } \sin
\Delta_{ 31 },$$
where $\Delta_{kj} = \Delta m_{kj}^{2} L /(4E)$ with $\Delta
m_{kj}^{2} = m_{k}^{2}-m_{j}^{2}$, $L$ is the baseline, $E$ is the
energy of neutrino, and $ J_{\alpha \beta} = {\cal I}m \left(
  U_{\alpha 1}U_{\alpha 2}^{*} U_{\beta 1}^{*} U_{\beta 2 } \right) =
\pm J_{CP}$, whereby it is called
Jarlskog-like invariant. The positive (negative) sign for
(anti-)cyclic permutation of the flavour indices $e$, $\mu$ and
$\tau$. For example for the oscillation between electron and muon
neutrinos, the transition probability $\nu_{\mu} \to \nu_{e}$ in
vacuum has the form
$P \left( \nu_{\mu} \to \nu_{e} \right)
 \simeq P_{\rm atm} + 2 \sqrt{P_{\rm atm} } \sqrt{ P_{\rm sol} }
 \cos \left( \Delta_{32} + \delta_{\rm CP} \right) + P_{\rm sol}$, where $\sqrt{P_{\rm
    atm}}=\sin\theta_{23}\sin2\theta_{13}\sin\Delta_{31}$ and
$\sqrt{P_{\rm sol}}=\cos\theta_{23} \cos \theta_{13}
\sin2\theta_{12}\sin\Delta_{21}$~\cite{Nunokawa2008338}. Hence, the
neutrino anti-neutrino asymmetry in the vacuum is
\begin{equation}\label{Eq:Asym:e_mu:1}
 A_{\mu e} =
  \frac{
   P \left( \nu_{\mu} \to \nu_{e} \right) - P \left( \bar{\nu}_{\mu} \to \bar{\nu}_{e} \right)
  }{
  P \left( \nu_{\mu} \to \nu_{e} \right) + P \left( \bar{\nu}_{\mu} \to \bar{\nu}_{e} \right)
  } =
  \frac{
   2 \sqrt{ P_{\rm atm} } \sqrt{ P_{\rm sol} } \sin \Delta_{32} \sin \delta_{\rm CP}
  }{
   P_{\rm atm} + 2 \sqrt{ P_{\rm atm} } \sqrt{ P_{\rm sol} } \cos \Delta_{32} \cos \delta_{\rm CP} + P_{\rm sol}}\,.
\end{equation}
\begin{figure}[hptb]
 \begin{center}
  \begin{tabular}{ccc}
   \includegraphics[width=0.44\linewidth]{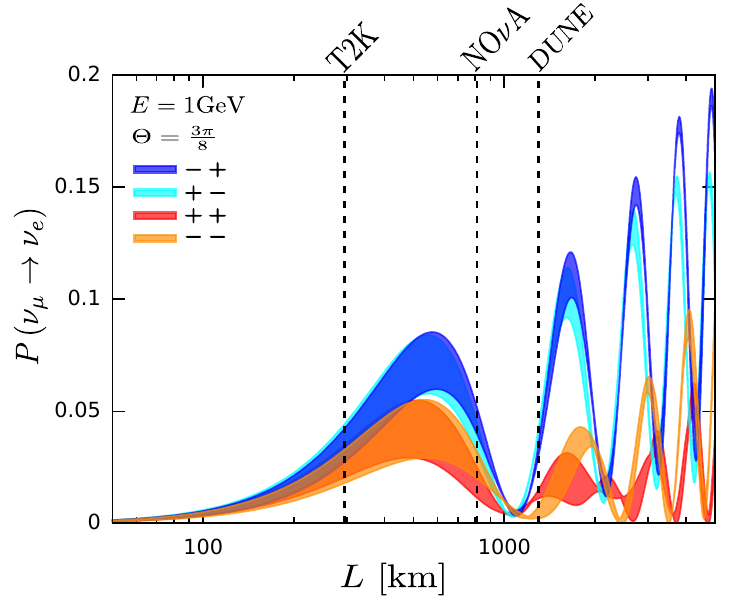} &
   \includegraphics[width=0.44\linewidth]{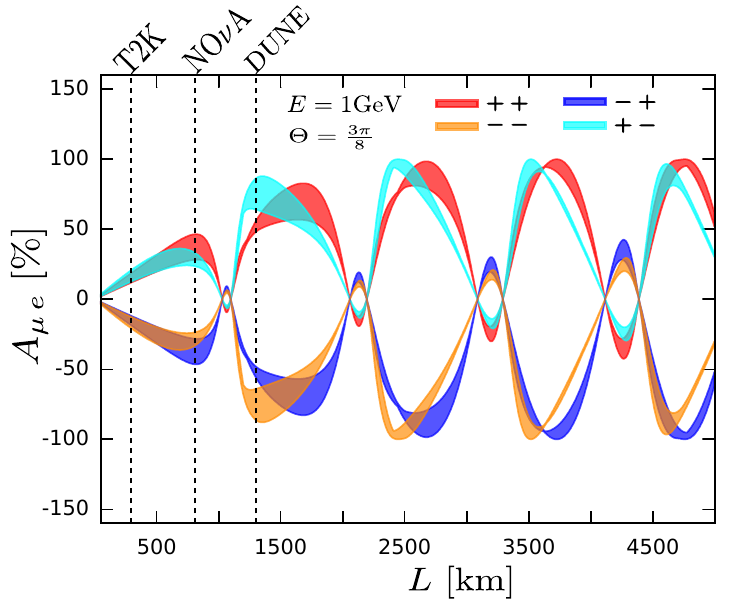}
  \end{tabular}
  \caption{ In the left panel we show the $\nu_{\mu} \to \nu_{e}$
    transition probability in matter for a neutrino energy of $E =
    1$GeV.  The right panel shows the neutrino anti-neutrino asymmetry
    ${\cal A}_{\mu e}$ in matter. The mixing angle $\theta_{23}$ is
    taken within the currently allowed 3$\sigma$ range $0.393 \leq
    \sin^{2} \theta_{23} \leq 0.643$~\cite{Forero:2014bxa}. The
    remaining neutrino oscillation parameters are fixed at their best
    fit values: $\Delta m_{21}^{2} = 7.60 \times
    10^{-5}\textrm{eV}^{2}$, $| \Delta m_{31}^{2} | = 2.48 \times
    10^{-3}\textrm{eV}^{2}$, $\sin \theta_{12} = 0.323$ and $\sin
    \theta_{13} = 0.0226$. The $\Theta$ parameter is fixed to the
    value $3\pi/8$. The figure corresponds to the case of normal
    ordering and the sign combinations refer to
    Eqs.~(\ref{Eq:P_atm-P_sol-Matter}) and
    (\ref{Eq:Asym:e_mu:2}). }\label{Fig:Delta_mu_e}
 \end{center}
\end{figure}
\begin{figure}[hptb]
 \begin{center}
  \begin{tabular}{cc}
   \includegraphics[width=0.44\linewidth]{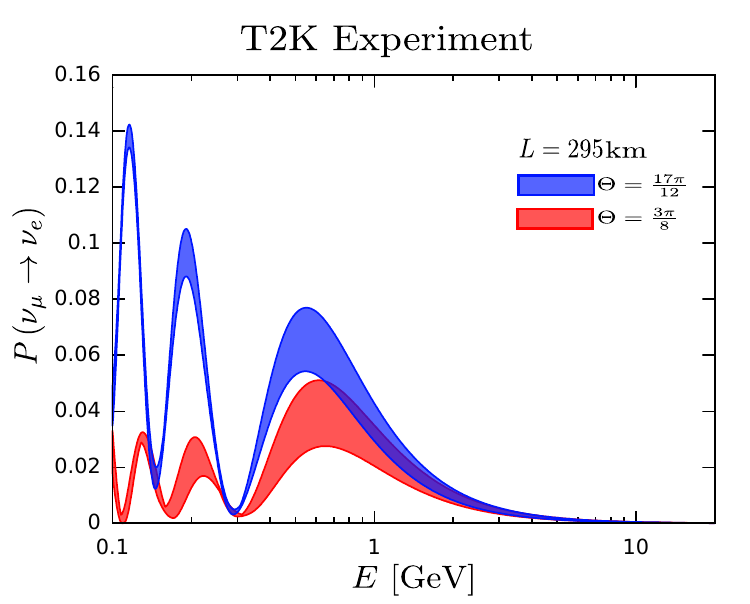} &
   \includegraphics[width=0.44\linewidth]{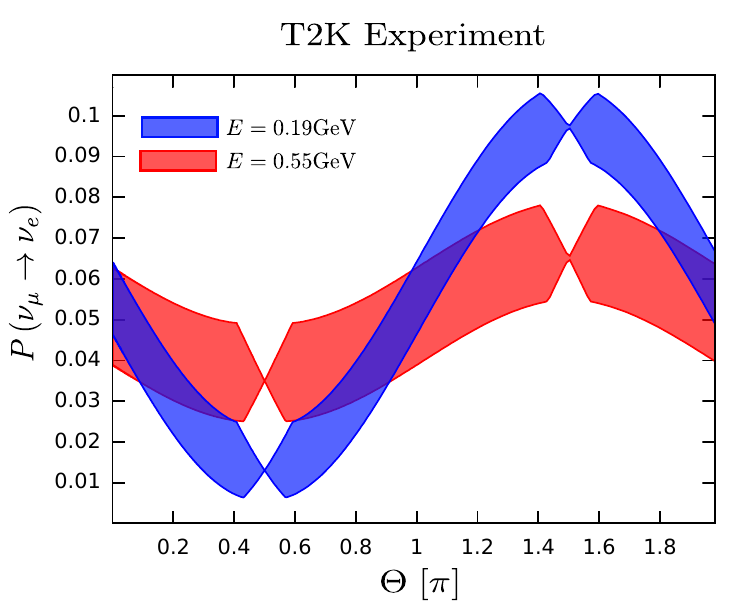}
  \end{tabular}
  \caption{The transition probability $P \left({\nu_{\mu}\to \nu_{e}}
    \right)$ at a baseline of 295km which corresponds to the T2K
    experiment. The mixing angle $\theta_{23}$ is taken within its
    currently allowed 3$\sigma$ regions $0.393 \leq \sin^{2}
    \theta_{23} \leq 0.643$~\cite{Forero:2014bxa}. Remaining
  oscillation parameters as in Fig.~\ref{Fig:Delta_mu_e} }\label{Fig:Exp:T2K}
 \end{center}
\end{figure}
\begin{figure}[hptb]
\begin{center}
\begin{tabular}{cc}
\includegraphics[width=0.44\linewidth]{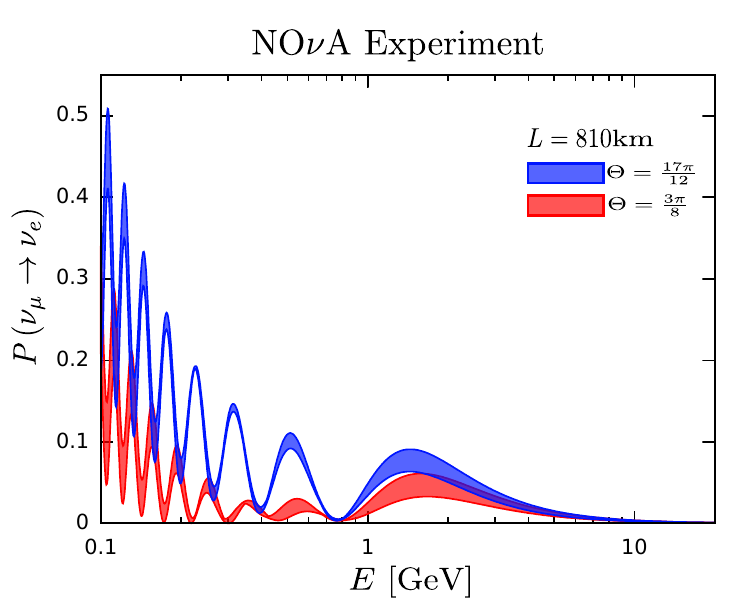} &
\includegraphics[width=0.44\linewidth]{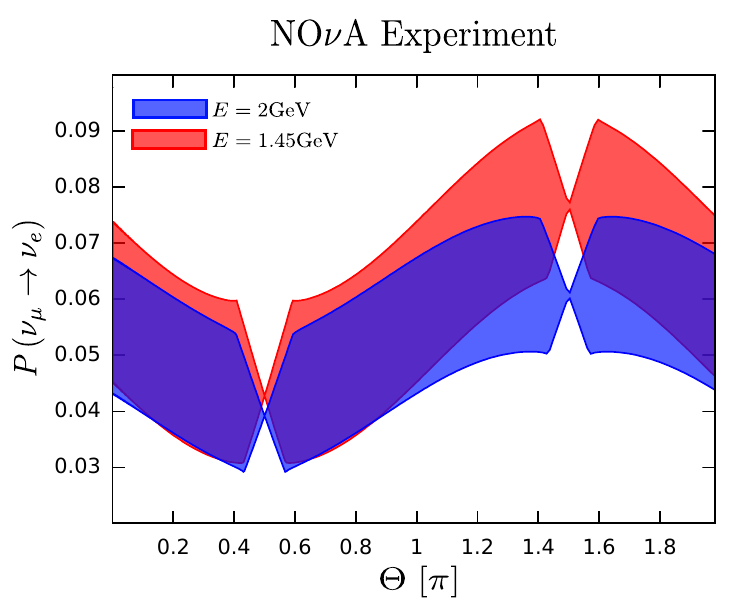}
\end{tabular}
\caption{ The transition probability $P \left({\nu_{\mu}\to \nu_{e}}
  \right)$ at a baseline of 810km which corresponds to the NO$\nu$A
  experiment. The mixing angle $\theta_{23}$ is considering into the
  currently allowed 3$\sigma$ regions $0.393 \leq
  \sin^{2}\theta_{23}\leq 0.643$~\cite{Forero:2014bxa}. Remaining
  oscillation parameters as in Fig.~\ref{Fig:Delta_mu_e}. }\label{Fig:Exp:NOvA}
\end{center}
\end{figure}
\begin{figure}[hptb]
\begin{center}
\begin{tabular}{cc}
\includegraphics[width=0.44\linewidth]{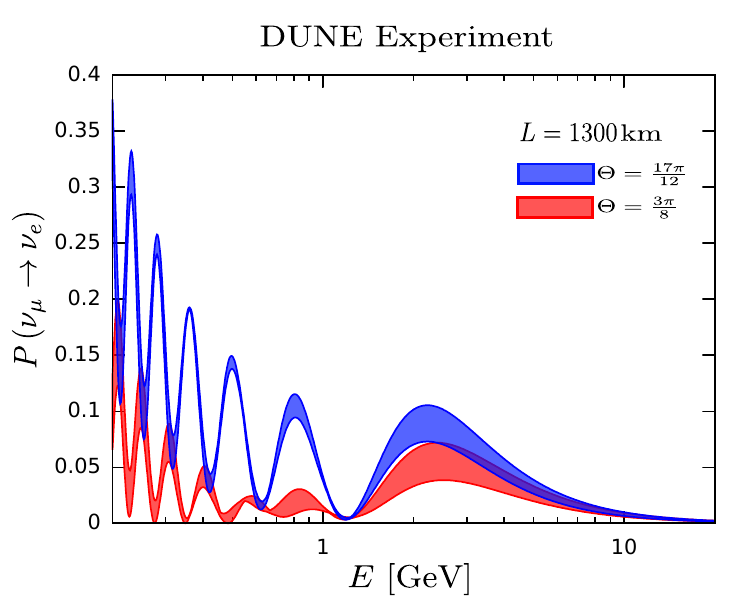} &
\includegraphics[width=0.44\linewidth]{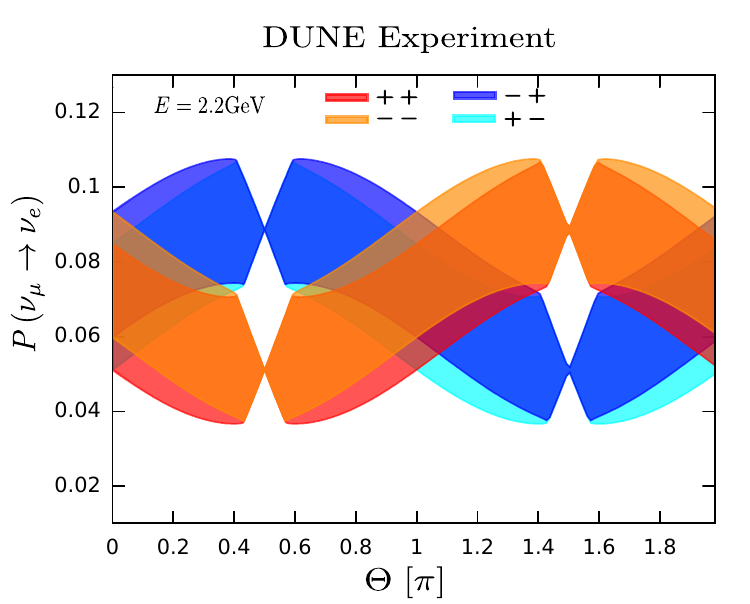}
\end{tabular}
\caption{ The transition probability $P \left( {\nu_{\mu}\to
      \nu_{e}}\right)$ at a baseline of 1300km, which corresponds to
  the DUNE proposal. The mixing angle $\theta_{23}$ is taken within
  the currently allowed 3$\sigma$ regions $0.393 \leq \sin^{2}
  \theta_{23} \leq0.643$~\cite{Forero:2014bxa}, while the remaining
  oscillation parameters are chosen as in Fig.~\ref{Fig:Delta_mu_e}. }\label{Fig:Exp:DUNE}
\end{center}
\end{figure}
In order to describe long baseline neutrino oscillations it is
important to include the effect of matter associated to neutrino
propagation in the Earth , as it can induce a fake CP violating
effect. In this case the expressions for $\sqrt{ P_{\rm atm} }$ and
$\sqrt{ P_{\rm sol} }$ in matter have the form:
\begin{equation}
 \begin{array}{l}\vspace{2mm}
  \sqrt{ P_{\rm atm} } =
  \sin \theta_{23} \sin 2\theta_{13}
  \frac{ \sin \left( \Delta_{31} - a L \right) }{\left( \Delta_{31} - a L \right) } \, \Delta_{31}\, , \quad
  \sqrt{ P_{\rm sol} } = \cos \theta_{23} \sin 2\theta_{12} \frac{ \sin(aL) }{ aL } \, \Delta_{21} \, ,
 \end{array}
\end{equation}
where $a = G_{F} N_{e}/\sqrt{2}$, $G_{F}$ is the Fermi constant and
$N_{e}$ is the density of electrons. The approximate value of $a$ is
$(3500 \text{km} )^{-1}$ for $\rho Y_{e} = 3.0\textrm{g\,cm}^{-3}$,
where $Y_{e}$ is the electron fraction~\cite{Nunokawa2008338}.  The
relative phase $({\Delta_{32}+\delta_{\rm CP}})$ between $\sqrt{P_{\rm
      atm}}$ and $\sqrt{P_{\rm sol}}$ remains unchanged.

  Within the framework of our generalized of $\mu - \tau$ reflection
  scenario, the transition probability $P(\nu_{\mu} \to \nu_{e})$ in
  matter has the form
\begin{equation}\label{Eq:P_atm-P_sol-Matter}
  P \left( \nu_{\mu} \to \nu_{e} \right) \simeq
  P_{\rm atm} + P_{\rm sol} \pm 2 \sqrt{ P_{\rm atm} } \sqrt{ P_{\rm sol} }
  \cos \left( {\Delta_{32}} \pm
  \arcsin \left( \frac{ \sin \Theta }{ \sin 2\theta_{23} } \right) \right) \, .
\end{equation}
The neutrino anti-neutrino asymmetry in matter is given by
\begin{equation}\label{Eq:Asym:e_mu:2}
 A_{\mu e} = \pm
  \frac{
   2 \sqrt{ P_{\rm atm} } \sqrt{ P_{\rm sol} } \sin \Delta_{23} \sin\Theta
  }{
   \left( P_{\rm atm}  + P_{\rm sol} \right) \sin 2\theta_{23}\pm 2 \sqrt{ P_{\rm atm} }
   \sqrt{ P_{\rm sol} }\sqrt{\sin^{2} 2\theta_{23} - \sin^{2} \Theta  } \; \cos \Delta_{23} \,
  },
\end{equation}
where $\sqrt{ P_{\rm atm} }$ and $\sqrt{ P_{\rm sol} }$ are given in Eq.~\eqref{Eq:P_atm-P_sol-Matter}.

In Fig.~\ref{Fig:Delta_mu_e} we show the $\nu_{\mu} \to \nu_{e}$
transition probability and the neutrino anti- neutrino asymmetry in
matter. In this figure we take the atmospheric mixing angle within its
currently allowed 3$\sigma$ region, while for the remaining neutrino
oscillation parameters are taken at their best fit
values~\cite{Forero:2014bxa}. In Figs.~\ref{Fig:Exp:T2K},
\ref{Fig:Exp:NOvA} we show the behavior of the transition probability
$P \left({\nu_{\mu}\to\nu_{e}} \right)$ in terms of neutrino energy
$E$ and the CP parameters $\Theta$ describing our approach, for
baseline values 295 and 810 km, which correspond to the current T2K
and NO$\nu$A experiments, respectively.

Note that so far we have discussed the predictions of our scenario for
neutrino oscillations at the T2K and NO$\nu$A experiments, for a fixed
sign combination in Eq.~(\ref{Eq:P_atm-P_sol-Matter}), which is
$(+,+)$.  We now consider the variation of our prediction with respect
to the choice of sign conbination. For definiteness we now consider
the future DUNE experiment. Fist we display in the left panel of
Fig.~\ref{Fig:Exp:DUNE} the behaviour of the $\nu_{\mu} \to \nu_{e}$
transition probability with respect to energy for the $(+,+)$ case and
two fixed values of the model parameter $\Theta$. In the right panel
of Fig.~\ref{Fig:Exp:DUNE} we display the model-dependence of the
$\nu_{\mu} \to \nu_{e}$ transition probability for different sign
combinations.

%-----------------------------------------------------------------------------
\section{Conclusion}\label{sec:conclusion}
%-----------------------------------------------------------------------------
CP violation is the least studied aspect of the lepton mixing matrix.
Other unknown features in the neutrino sector include the neutrino
mass ordering and the octant of the atmospheric mixing parameter
$\theta_{23}$, not yet reliably determined by current global
oscillation fits. In this letter we have proposed a generalized
$\mu-\tau$ reflection scenario for leptonic CP violation and derived
the corresponding restrictions on lepton flavor mixing parameters. We
found that the ``Majorana'' phases are predicted to lie at their
CP-conserving values with important implications for the \onbb
amplitudes, which we work out in detail. In addition to this
prediction concerning the vanishing of the ``Majorana-type'' CP
violation, we have obtained a new correlation between the atmospheric
mixing angle $\theta_{23}$ and the ``Dirac'' CP phase $\delta_{\rm
  CP}$.  Only in a very specific limit our CP transformation reduces
to standard $\mu-\tau$ reflection, for which $\theta_{23}$ and
$\delta_{\rm CP}$ become both maximal. We have also analysed the
phenomenological implications of our scheme for present as well as
upcoming neutrino oscillation experiments T2K, NO$\nu$A and DUNE.
In analogy to the case of $\mu-\tau$ reflection symmetry, we expect that in our generalized $\mu-\tau$ reflection symmetry approach it may be possible to predict the value of the angle $\Theta$. This may arise from some particular residual flavor symmetries which close, say, to a finite group~\cite{Joshipura:2015dsa}, or in the context of a flavor symmetry combined with the generalized CP symmetry~\cite{Feruglio:2012cw}. Detailed study of this possibility is left for future work.

%%%%%%%%%%%%%%%%%%%
%	A C K N O W L E D G M E N T S
%%%%%%%%%%%%%%%%%%%
\begin{acknowledgments}
%%%%%%%%%%%%%%%%%%%%%
  This work was supported by the National Natural Science Foundation
  of China under Grant Nos. 11275188, 11179007 and 11522546 (P.C. and
  G.J.D.); by the Spanish grants FPA2014-58183-P, Multidark
  CSD2009-00064 and SEV-2014-0398 (MINECO), and PROMETEOII/2014/084
  (Generalitat Valenciana) (J.W.F.V. and F.G.C.), and the Mexican
  grants \textit{CONACYT} 250950, 132059, and PAPIIT IN111115
  (F.G.C.). We thank M. Tortola for helpful discussions.
\end{acknowledgments}

%\bibliographystyle{bib_style_T1}
%\bibliography{newrefs,merged_Valle,param,newrefs_ding}

\end{document}